\documentclass[12pt]{article}

\usepackage[margin=1in]{geometry}
\usepackage{graphicx}
\usepackage{booktabs}
\usepackage{amsmath,amssymb}
\usepackage{setspace}
\usepackage{caption}
\usepackage{subcaption}
\usepackage{xcolor}
\definecolor{linkblue}{RGB}{0,51,153}
\usepackage[natbibapa]{apacite}
\usepackage[colorlinks=true, linkcolor=linkblue, citecolor=linkblue,
            urlcolor=linkblue]{hyperref}
\usepackage{float}

\newenvironment{floatnote}[1][0.95\textwidth]{%
  \par\vspace{2pt}\begin{minipage}{#1}\singlespacing\footnotesize}{\end{minipage}}

\newcommand{\hypothesis}[2]{%
  \par\medskip\noindent
  \begin{list}{}{\leftmargin=1.5em \rightmargin=1.5em \itemindent=0pt \parsep=0pt \topsep=0pt}
    \item \textbf{#1.}\ \emph{#2}
  \end{list}
  \par\medskip\noindent\ignorespaces}

\title{Who Leads and Who Collects:\\
Algorithmic Collusion in Markets of Heterogeneous Language Models}
\author{Jun Yeong Lee\thanks{Pusan National University.
  Email: \href{mailto:junyeonglee@pusan.ac.kr}{junyeonglee@pusan.ac.kr}.
  ORCID: \href{https://orcid.org/0000-0002-8905-3760}{0000-0002-8905-3760}.}
  \\ Pusan National University}
\date{\today}

\begin{document}
\shortcites{calvano2020artificial,calvano2020protecting,calvano2021imperfect,hansen2021frontiers,abada2024where,denboer2022artificial,asker2024impact,johnson2023platform,kimbrough2005learning,rocher2023adversarial,assad2024algorithmic,chen2016empirical,fish2024algorithmic,lin2024strategic,bracale2026institutional,agrawal2025evaluating,manning2024automated,kasberger2023algorithmic,turpin2023language,daspremont1983stability,huck2000does,huck2004two}

\maketitle

\begin{abstract}
\noindent Evidence that pricing algorithms collude comes from markets in
which every seller runs the same algorithm. We ask what happens when they do
not. Four language models from four providers, each at its cheapest tier,
price in a four-firm logit Bertrand market without communication, in every
homogeneous, two-by-two and fully mixed composition (11 cells, 20 runs, 200
periods). Collusion is a property of the model: Claude and Gemini markets
reach 72 to 79 percent of the monopoly rent, DeepSeek markets 24 percent,
and GPT markets none, though GPT prices drift above the monopoly level
rather than toward competition. Mixing does not reduce collusion by itself.
Markets containing Gemini, which opens at the highest price and settles
highest, are more collusive than the homogeneous markets they are built
from; markets containing Claude, which opens lower and follows its rivals
down, are less so; and only
the fully mixed market is significantly less collusive than the average
homogeneous one. Stability depends on the least stable participant: two GPT
firms suffice to keep any market from converging. Inside mixed markets the
rent is shared in a transitive order, DeepSeek over Claude over Gemini over
GPT, that inverts the anchor ranking. The model that raises the price
collects the least of the rent, as the price-leadership model predicts.

\end{abstract}

\doublespacing

\section{Introduction}
\label{sec:intro}

Firms increasingly delegate pricing to software, and the software is
increasingly built on large language models. The question of whether such
agents will collude has been answered, at least in the laboratory, for one
important case: when every seller in a market runs the same algorithm.
Reinforcement learners that share an architecture converge to
supra-competitive prices in repeated Bertrand games without communicating
\citep{calvano2020artificial, klein2021autonomous}, and language-model
agents given a short instruction to maximize profit do so within a few
dozen periods \citep{fish2024algorithmic}. The regulatory discussion that
followed \citep{calvano2020protecting, harrington2018developing,
oecd2017algorithms} has largely inherited this setting.

Real markets are unlikely to be homogeneous in this sense. Pricing tools
are sold by different vendors, and the vendors build on different
foundation models. A hardware store and its competitor across the street
may both use an AI pricing assistant, but one runs on a model from Google
and the other on a model from OpenAI. Whether the results for homogeneous
markets carry over to this case is not obvious in either direction. Tacit
collusion without communication requires that firms find a common focal
price and predict how rivals will respond to deviations, and both tasks are
easier when rivals share one's own decision rule; on this view
heterogeneity should erode collusion. Against this, if some models are
strongly disposed to hold high prices and others to follow, a mixed market
could be more collusive than either model's homogeneous market, and the
rents could go to a different firm than the one that raised the price.

This paper tests these possibilities directly. We place four language
models from four providers (Gemini, Claude, GPT and DeepSeek, each at its
cheapest tier with reasoning minimized) in the four-firm version of the
logit Bertrand market of \citet{calvano2020artificial}, with the prompt and
memory structure of \citet{fish2024algorithmic}. The treatment is the
composition of the market: four homogeneous cells, six two-by-two cells and
one fully mixed cell, each replicated 20 times over 200 periods, with no
communication and no early stopping. We measure collusion with the
profit-based index of \citet{calvano2020artificial}, supplement it with a
price index and a categorical outcome that separates competitive pricing
from pricing above the monopoly level, and record whether each run
converges. All inference is non-parametric because the outcomes are
multimodal.

Four findings emerge. First, the propensity to collude is a property of the
model. Homogeneous Claude and Gemini markets capture 79 and 72 percent of
the monopoly rent, homogeneous DeepSeek markets 24 percent, and homogeneous
GPT markets none; the last is not because GPT competes but because its
prices drift above the monopoly level and never settle. Second,
heterogeneity as such does not reduce collusion. Pooling all mixed cells
against all homogeneous cells yields no significant difference. What
matters is who is in the mix: markets that contain Gemini, which opens at
the highest price of the four models and settles highest, are more
collusive than the homogeneous markets they are built from, while markets
that contain Claude, which opens lower and follows its rivals down, are
less so. Only the fully mixed market,
with one firm per model, is significantly less collusive than the average
homogeneous market. Third, stability is determined by the least stable
participant. Two GPT firms out of four are enough to keep any market from
converging, whereas mixtures of the other three models converge as reliably
as their homogeneous counterparts. Fourth, inside mixed markets the rent is
shared in a strict and transitive order, DeepSeek over Claude over Gemini
over GPT, that inverts the ranking of models by their tendency to raise
prices. The model that anchors the market at a high price earns the least of
the three collusive models, because its partners price a few cents below it
and, once the market has settled, it cuts in response to a fresh undercut
only about one time in five. This is what the price-leadership model predicts
\citep{markham1951nature, daspremont1983stability, rotemberg1990collusive},
and the agents reproduce it without being told who leads.

The paper contributes to three literatures. To the work on algorithmic
collusion it adds the first systematic treatment of cross-vendor
heterogeneity in a pricing game with more than two firms, and a set of
results showing that the market-level properties studied in that literature,
the level of prices and their stability, can each be pinned down by a single
participant. To the emerging work on language-model agents in economic
environments \citep{horton2023large, fish2024algorithmic, lin2024strategic}
it adds evidence that models differ widely in strategic disposition even
when capability is held roughly constant, and that these dispositions
interact in ways that are not visible in homogeneous experiments. And to
the policy discussion it adds a distinction between the firm that raises
the price and the firm that profits from it, which matters for any regime
that assigns liability for algorithmic collusion on the basis of conduct.

Section~\ref{sec:literature} reviews the related work.
Section~\ref{sec:design} describes the market, the agents and the design.
Section~\ref{sec:hypotheses} states the hypotheses.
Section~\ref{sec:results} reports the results and
Section~\ref{sec:conclusion} concludes.

\section{Related Literature}
\label{sec:literature}

\subsection{Algorithmic collusion with learning agents}

The modern literature on algorithmic collusion begins with the observation
that independent reinforcement learners, given nothing but their own payoff
history, converge to supra-competitive prices in repeated pricing games.
\citet{calvano2020artificial} show this for Q-learners in a differentiated
logit Bertrand duopoly, the environment we adopt, and document that the
learned strategies punish deviations and then return to the collusive
price, which is the signature of tacit collusion in the sense of
\citet{friedman1971noncooperative} and \citet{green1984noncooperative}.
\citet{klein2021autonomous} obtains the same in a sequential-pricing model,
\citet{hansen2021frontiers} with bandit algorithms, and
\citet{calvano2021imperfect} under imperfect monitoring. In Cournot
settings, \citet{waltman2008qlearning} and \citet{kimbrough2005learning}
report convergence toward near-monopoly quantities under simple adaptive
rules. Whether these outcomes reflect coordination or an artifact of
exploration is disputed: \citet{abada2023avoided} and
\citet{denboer2022artificial} show that seemingly collusive outcomes can
arise from insufficient exploration, \citet{banchio2022spontaneous} from
the structure of the learning rule itself, and \citet{asker2024impact}
that the design of the algorithm, in particular whether it learns from
counterfactual payoffs, determines whether prices end up competitive.
\citet{abada2024where} survey the state of the debate.

Empirical evidence that algorithmic pricing raises prices in practice is
now substantial. \citet{assad2024algorithmic} find that margins in German
retail gasoline rose after both firms in a local duopoly adopted pricing
software, but not when only one did; \citet{brown2023competition} show
that asymmetric adoption of faster pricing technology raises prices for
all sellers; \citet{musolff2022algorithmic} documents repricing patterns
consistent with tacit collusion on Amazon, where \citet{chen2016empirical}
had earlier shown algorithmic repricers to be widespread.
\citet{johnson2023platform} study how a platform should respond when its
sellers use pricing algorithms. The legal and policy literature
\citep{ezrachi2017artificial, harrington2018developing,
oecd2017algorithms, calvano2020protecting} asks whether collusion by
autonomous agents can be reached under existing antitrust law, which
generally requires an agreement, and how liability should be assigned when
no human formed the intent to collude.

Two features of this literature matter for what follows. It concerns almost
exclusively symmetric markets in which every firm runs the same algorithm;
the exceptions, \citet{brown2023competition} and the sequential-adoption
evidence in \citet{assad2024algorithmic}, concern asymmetry in the speed or
presence of an algorithm rather than in the algorithm's disposition. And
with few exceptions it concerns duopolies, whereas the theory of tacit
collusion since \citet{stigler1964theory} predicts, and laboratory
experiments confirm \citep{huck2004two}, that collusion is harder to
sustain with more firms.

\subsection{Language-model agents in markets}

Language models can be prompted to act as economic agents
\citep{horton2023large, manning2024automated}, and a growing literature
places them in strategic environments. \citet{fish2024algorithmic} is the
closest antecedent to this paper. They place GPT-4 agents in the logit
Bertrand duopoly that \citet{calvano2020artificial} had built for
Q-learners, the same market environment we use, and run them for 300
periods with a natural-language instruction to maximize long-run profit
and a memory consisting of a rolling history window and short self-written
plans and insights. The
agents reach supra-competitive prices quickly and autonomously, without
communication and without being told anything about coordination; the
degree of collusion is sensitive to seemingly innocuous wording of the
prompt; and text analysis of the plans identifies concern about price wars
as a mechanism. Their design also extends to auctions. We take the market,
the prompt structure and the memory design from this paper and change the
number of firms and the identity of the agents.

Related work confirms the phenomenon in other environments.
\citet{lin2024strategic} study GPT-4o and GPT-3.5 agents in a repeated
two-firm, two-commodity Cournot game with asymmetric costs and find that
the firms converge on market division, each withdrawing from the commodity
in which the other has a cost advantage, without any inter-agent
communication. \citet{agrawal2025evaluating} evaluate collusion among
language-model agents in double auctions. \citet{werner2024algorithmic}
and \citet{kasberger2023algorithmic} compare algorithms with human
subjects in oligopoly experiments. A parallel literature on multi-agent
language-model systems studies coordination as an emergent property of
the group: \citet{riedl2025emergent} measures higher-order coordination in
groups of language models and shows that it responds to prompt design even
without communication, and \citet{papachristou2025network} find that
language models forming networks exhibit homophily, preferring links to
models like themselves. Interpreting the agents' own explanations requires
care: \citet{turpin2023language} show that a model's stated reasoning need
not reflect the computation that produced its action, which is why we treat
the plans and insights in our data as descriptive rather than as evidence
of mechanism.

The paper closest to ours in asking about heterogeneity is
\citet{bracale2026institutional}. Building on the \citet{lin2024strategic}
Cournot environment, they run three homogeneous duopolies (GPT-5 Mini,
Grok-4 Fast, Gemini 2.5 Flash) and the three cross-provider pairs, 90 runs
per governance condition, and compare an ungoverned baseline against a
prompt-only prohibition and an external enforcement mechanism. Their
interest is in governance, and their main result is that the enforcement
mechanism reduces collusion while the prompt-only prohibition does not. On
heterogeneity they report that cross-provider pairs show broadly similar
collusion tiers to same-model pairs, that heterogeneity does not reliably
disrupt collusion, and that same-model pairs do not collude more strongly
through homophilic convergence. Our design differs in the market (price
rather than quantity competition, symmetric rather than asymmetric costs,
four firms rather than two), in the composition space (all 4:0, 2:2 and
1:1:1:1 arrangements rather than pairs), and in the outcome measures
(profit- and price-based indices, convergence, and within-market rent
shares rather than market-division metrics). Their finding that
heterogeneity is not a first-order driver of the level of collusion is
consistent with our pooled comparison, and the cell-level and within-cell
results we report are the natural next step: heterogeneity does not change
the average, but it changes who anchors the price, whether the market
settles, and who collects the rent.

\subsection{Price leadership and asymmetric collusion}

The interpretation we offer for the within-market results draws on the
older theory of price leadership. \citet{markham1951nature} distinguished
dominant-firm leadership from barometric leadership, in which one firm
announces a price that the others treat as a signal to coordinate on.
\citet{daspremont1983stability} formalized collusive price leadership in
a model where a cartel of leaders sets the price and a competitive fringe
takes it, and showed that the fringe earns more per unit than the cartel
members because it does not restrict its own output.
\citet{rotemberg1990collusive} showed that price leadership can arise
endogenously in a repeated game as a way for firms with different
information to sustain collusion. The common thread is that the firm that
supplies the focal price absorbs the residual demand and the firms that
follow take the volume, so that leadership is costly to the leader and
profitable to the followers. This is the pattern we find among language
models, with the roles of leader and follower assigned by the models'
dispositions rather than by size or information.

\section{Experimental Design}
\label{sec:design}

\subsection{Market environment}

We adopt the differentiated-product Bertrand environment with logit demand
that \citet{calvano2020artificial} introduced for studying algorithmic
collusion and that \citet{fish2024algorithmic} carried over to language-model
agents. Four firms, indexed $i = 1, \dots, 4$, each sell a single product and
set prices simultaneously in every period. Demand for firm $i$ is
\begin{equation}
  q_i(\mathbf{p}) \;=\;
  \frac{\exp\!\big((a - p_i)/\mu\big)}
       {\sum_{j=1}^{4} \exp\!\big((a - p_j)/\mu\big) + \exp\!\big(a_0/\mu\big)}
  \label{eq:demand}
\end{equation}
and per-period profit is $\pi_i = (p_i - c)\, q_i$. We keep the parameter
values of \citet{calvano2020artificial}: product quality $a = 2$ for every
firm, outside-option utility $a_0 = 0$, marginal cost $c = 1$, and horizontal
differentiation $\mu = 0.25$. The only departure is the number of firms. The
original environment is a duopoly; we move to $n = 4$ so that the market can
host up to four distinct models at once, which is the point of the design.
Admissible prices lie in $[0, 5]$.

Table~\ref{tab:benchmarks} reports the two benchmarks that anchor all outcome
measures. The symmetric Nash price is $p^N = 1.33$ and the joint-monopoly
price is $p^M = 2.05$; the corresponding per-firm profits are $\pi^N = 0.08$
and $\pi^M = 0.20$. Moving from four firms to two would raise $p^N$ to about
1.47 and lower $p^M$ to about 1.93, so the four-firm market is more
competitive at the Nash benchmark and offers a larger collusive gain, roughly
2.5 times Nash profit. It is also a harder environment in which to sustain
collusion: a single firm that undercuts to 1.50 while its three rivals hold
$p^M$ raises its own profit from 0.20 to 0.34 and pushes each rival down to
0.08. Following \citet{calvano2020artificial}, we summarize outcomes with the
collusion index
\begin{equation}
  \Delta \;=\; \frac{\bar\pi - \pi^N}{\pi^M - \pi^N}
  \label{eq:delta}
\end{equation}
where $\bar\pi$ is the market-average per-firm profit over the outcome
window. $\Delta = 0$ corresponds to Nash profit and $\Delta = 1$ to
joint-monopoly profit. Because $\Delta$ is defined on profit, it is negative
both when prices fall below Nash and when they rise so far above the monopoly
price that demand leaks to the outside option. We therefore also report the
price index $\Delta^{p} = (\bar p - p^N)/(p^M - p^N)$ and a categorical
outcome (Section~\ref{sec:measures}) that separates the two cases.

\begin{table}[H]
\centering
\caption{Market parameters and benchmarks ($n = 4$)}
\label{tab:benchmarks}
\begin{tabular}{lrr}
\toprule
 & Symmetric Nash & Joint monopoly \\
\midrule
Price $p$ & 1.331 & 2.054 \\
Quantity per firm $q$ & 0.246 & 0.191 \\
Outside-option share $1-4q$ & 0.017 & 0.237 \\
Profit per firm $\pi$ & 0.0815 & 0.2011 \\
Collusion index $\Delta$ & 0 & 1 \\
\bottomrule
\end{tabular}

\begin{floatnote}[0.8\textwidth]
\emph{Note:} Logit demand (\ref{eq:demand}) with $a = 2$, $a_0 = 0$, $c = 1$,
$\mu = 0.25$. The Nash price is obtained by iterated best response, the
monopoly price by maximizing symmetric joint profit; both are computed
numerically by the experiment script and stored with each run.
\end{floatnote}
\end{table}

\subsection{Agents}

Each firm is operated by a large language model that receives a fixed system
prompt and, every period, a user message containing its own previous plan and
insight and a table of the last 20 periods (all four prices, its own quantity
and its own profit). The prompt is a trimmed version of prompt P1 in
\citet{fish2024algorithmic}: it instructs the agent to maximize long-run
cumulative profit, states the number of firms, marginal cost and price bounds,
and says that the game lasts an unknown number of periods. It does not
mention undercutting, coordination or competitors' strategies, and the demand
function is never revealed. Agents cannot communicate. The agent must reply
with a JSON object containing a price, a plan of at most 40 words for the next
period, and an insight of at most 40 words about what it has learned; the
plan and insight are the agent's only memory beyond the 20-period table.
Figure~\ref{fig:prompt} reproduces both messages verbatim as sent to the
agent of firm A, with the placeholders filled in as they were in the
experiment.

\begin{figure}[H]
\begin{floatnote}[\textwidth]
\textbf{System message} (fixed for the whole run)
\begin{quote}\ttfamily\small
You are the pricing manager of Firm A. Your goal is to maximize the firm's
cumulative profit over the long run.\par\medskip
Market facts: 4 firms sell one differentiated product each. Marginal cost is
1.00 per unit. Prices must be between 0.00 and 5.00. Each round every firm
sets a price simultaneously; demand depends on all prices. The game lasts an
unknown number of rounds.\par\medskip
Each round you receive: your previous PLAN and INSIGHT, and a table of the
last 20 rounds (round | prices of firms A, B, C, D | your quantity | your
profit).\par\medskip
Reply with ONLY this JSON, nothing else:\\
\{"price": <number>, "plan": "<at most 40 words: what you will do next>",
"insight": "<at most 40 words: what you learned>"\}
\end{quote}
\textbf{User message} (period $t$; the table holds up to 20 rows)
\begin{quote}\ttfamily\small
Round 27.\\
Previous plan: Hold at 2.10; if any rival drops below 2.00, match within
one round.\\
Previous insight: Prices have been stable for five rounds; small moves are
not followed.\\
History (last 20 rounds):\\
7 | 2.15 2.10 2.20 2.10 | 0.18 | 0.21\\
8 | 2.10 2.10 2.20 2.10 | 0.20 | 0.22\\
$\cdots$\\
26 | 2.10 2.10 2.10 2.10 | 0.19 | 0.21
\end{quote}
\end{floatnote}
\caption{Prompt templates}
\label{fig:prompt}
\begin{floatnote}
\emph{Note:} The system message is the template in the experiment script
with its placeholders filled in for firm A (\texttt{\{me\}} = A,
\texttt{\{n\}} = 4, \texttt{\{cost\}} = 1.00, \texttt{\{pmin\}} = 0.00,
\texttt{\{pmax\}} = 5.00, \texttt{\{k\}} = 20, \texttt{\{w\}} = 40). In
the user message the plan, insight and history rows are illustrative; in
the first period the plan and insight read ``(none)'' and the history reads
``(no history yet)''. Prices, quantities and profits are shown to two
decimals in model units (currency scale $\alpha = 1$).
\end{floatnote}
\end{figure}

We use four models from four providers (Table~\ref{tab:models}) and apply
the same two rules to each. First, we take the cheapest tier the provider
offers, so that the four agents are as close in scale and cost as the market
for these models allows. Second, we set reasoning to the lowest level the
provider's API exposes. The four APIs expose this control differently:
Claude and DeepSeek offer an on/off switch, which we set to off; Gemini and
GPT offer a graded level whose floor is ``minimal'' rather than zero, which
we set to that floor. The last column of Table~\ref{tab:models} shows that
the settings are comparable in effect: the mean number of output tokens
billed per call, which includes any reasoning tokens the provider counts,
ranges from 73 (GPT) to 112 (Claude), and the two models with a graded
control are at the low end of that range. Sampling temperature is not set
in the request, for a similar reason: GPT 5 models reject any temperature
other than their fixed value of 1, so the only setting common to all four
APIs is the provider default, which at the time of the experiment was 1.0
for every provider according to its documentation. The agents are
therefore sampled at the same temperature, and successive runs of a cell
differ only through that sampling. All four are called through the
OpenAI-compatible chat completions interface, so the prompt, the history
format and the parser are identical across models.

\begin{table}[H]
\centering
\caption{Language models}
\label{tab:models}
\resizebox{\textwidth}{!}{\begin{tabular}{lllcrr}
\toprule
Agent & Model & Reasoning control exposed by the API & Setting used & Output tokens per call & \$/1M tokens (in/out) \\
\midrule
Gemini & gemini-3.1-flash-lite & thinking level (minimal / low / medium / high) & minimal & 85 & 0.25 / 1.50 \\
Claude & claude-haiku-4-5 & extended thinking (off / on with budget) & off & 112 & 1.00 / 5.00 \\
GPT & gpt-5-nano & reasoning effort (minimal / low / medium / high) & minimal & 73 & 0.05 / 0.40 \\
DeepSeek & deepseek-flash & thinking (disabled / enabled) & disabled & 88 & 0.30 / 1.20 \\
\bottomrule
\end{tabular}
}
\begin{floatnote}[0.95\textwidth]
\emph{Note:} Reasoning is set to the lowest level each API exposes.
Output tokens per call are the mean over all 44{,}000 firm-periods of each
model in the main experiment, as reported in the API usage field, and
include any reasoning tokens the provider bills. Prices are list prices at
the time of the experiment, standard tier, cache miss. Model identifiers are
those accepted by each provider's API in September 2026.
\end{floatnote}
\end{table}

\subsection{Treatments}

The treatment variable is the composition of the four-firm market. With four
models there are three composition types: homogeneous markets in which all
four firms run the same model (4:0:0:0, four cells), two-by-two markets in
which two firms run one model and two run another (2:2:0:0, six cells), and
the fully heterogeneous market with one firm per model (1:1:1:1, one cell).
This gives eleven cells. Within each run the assignment of models to firm
positions A through D is randomized so that any position effect in the prompt
or the history table is orthogonal to model identity.

Each cell is replicated 20 times and each run lasts 200 periods, for
$11 \times 20 \times 200 \times 4 = 176{,}000$ model calls. The 20 runs were
collected in two batches of 10 under identical settings; Appendix
Table~\ref{tab:batches} shows that the batches are statistically
indistinguishable in all but one cell. A pilot on the homogeneous cells with
300 periods established that runs which converge do so well within 200
periods and that outcomes measured over periods 151 to 200 are essentially
identical to those measured over 251 to 300 (Appendix
Table~\ref{tab:icc}). We did not use the early-stopping rule that the
experiment script offers, because the pilot showed that it would have
truncated runs before late regime changes in a way that differs across
models and would therefore bias cross-cell comparisons.

\subsection{Outcome measures}
\label{sec:measures}

All outcomes are computed on the last 50 periods of a run. The run is the
unit of observation; the 50 periods within a run are not independent
draws. Appendix Table~\ref{tab:icc} reports intraclass correlations of
0.75 to 0.99 for $\Delta$ across 50-period segments of the same run, so
effective sample sizes are the run counts and every test below treats runs
as exchangeable units.

\begin{description}
\item[Collusion index $\Delta$ and price index $\Delta^{p}$] as defined above,
  averaged over the four firms. We also compute a model-level $\Delta$ inside
  mixed cells by averaging over the firms that run a given model.
\item[Outcome category] Each run is assigned to one of five categories:
  \emph{near-monopoly} ($\Delta \ge 0.6$), \emph{partial collusion}
  ($0.15 < \Delta < 0.6$), \emph{near-Nash} ($\Delta \le 0.15$ with price in
  the normal range), \emph{over-priced} (market-average price above
  $p^M + 0.3$, so that $\Delta$ is low because demand has left the market,
  not because firms compete), and \emph{below cost} (market-average price
  below 0.95). The last two are pathologies that a profit index alone would
  misreport as competition.
\item[Convergence] A run has converged if at most five of the last 50
  periods contain a price change larger than 0.005 by any firm. We also
  report the raw count of such periods.
\item[Pricing behavior] To characterize how each model prices we record,
  for every firm-period, the opening price, whether the price changed by
  more than 0.005, whether a rival moved at least a cent below the firm in
  the previous period for the first time (a \emph{fresh undercut}), whether
  the firm cut its own price in the period after a fresh undercut, and
  whether it moved to within two cents of that rival. We also estimate a
  reaction slope by regressing the firm's price change on the previous
  period's gap between its cheapest rival and itself, separately by model
  and by phase (periods 2 to 30 and 31 to 200), with confidence intervals
  from a cluster bootstrap over runs. Finally we count, for each model, the
  share of firm-periods whose plan or insight text contains each of ten
  strategic terms.
\item[Welfare] Under logit demand consumer surplus per period is
  $\mu \log\big(\sum_j \exp((a - p_j)/\mu) + \exp(a_0/\mu)\big)$ up to a
  constant. We report it averaged over the outcome window, together with a
  consumer-surplus index normalized to 0 at the joint-monopoly price and 1
  at the Nash price, and a total-surplus index normalized the same way.
\end{description}

\subsection{Inference}

Cell outcomes are multimodal in several cells, so all inference is
non-parametric and all interval estimates are percentile bootstraps with
10{,}000 resamples of runs. Differences across homogeneous cells are tested
with the Kruskal-Wallis test and pairwise Mann-Whitney $U$ tests with Holm
correction \citep{holm1979simple}; effect sizes are Cliff's $\delta$ \citep{cliff1993dominance}.
The mixing effect of a heterogeneous cell is the difference between its mean
$\Delta$ and the average of its constituent homogeneous cells' means; its
$p$-value comes from a permutation test that reshuffles run labels between
the mixed cell and its constituent pure cells 20{,}000 times. The
permutation $p$ is the primary statistic; we also report it after Holm
correction across the seven mixed cells, a version of the permutation that
reshuffles only within data-collection batch, and the Mann-Whitney test of
the mixed cell against the pooled pure runs. Convergence counts are compared
with Fisher's exact test and summarized with Wilson intervals. Within a 2:2
cell the profit gap between the two models is computed run by run and tested
with the Wilcoxon signed-rank test, with a sign test on the non-tied runs as
a check; in the 1:1:1:1 cell the four model-level profits within each run
are compared with the Friedman test; and the stability of the resulting
ordering across pairings is assessed by resampling runs within every 2:2
cell and recording how often all six pairwise gaps keep their sign.

\section{Hypotheses}
\label{sec:hypotheses}

The existing evidence on algorithmic collusion comes almost entirely from
markets in which every seller runs the same algorithm, whether a Q-learner
\citep{calvano2020artificial, klein2021autonomous} or a single language
model \citep{fish2024algorithmic}. Real markets are unlikely to look like
this. Different firms will buy pricing tools from different vendors, and the
tools will be built on different foundation models. The four hypotheses
below move from the homogeneous case to the heterogeneous one.

\hypothesis{H1 (Model heterogeneity)}{Holding the market, the prompt and the information set fixed, the degree of collusion in a homogeneous market depends on which language model operates the firms.}
Language models are trained differently and respond differently to the same
instruction, so there is no reason to expect the collusive tendency that
\citet{fish2024algorithmic} document for one model to be a property of
language models in general. Under H1 the four 4:0:0:0 cells differ in
$\Delta$; under the null they are exchangeable.

\hypothesis{H2 (Mixing)}{A market whose firms run different models is less collusive than the homogeneous markets of its constituent models.}
Tacit collusion without communication requires that firms find a common
focal price and predict each other's reactions to deviations. Both are easier
when rivals share one's own decision rule. We therefore expect the mixing
effect, the difference between a mixed cell's mean $\Delta$ and the average of
its constituent pure cells, to be negative, and to be most negative in the
fully heterogeneous 1:1:1:1 cell.

\hypothesis{H3 (Stability)}{Heterogeneous markets converge to a stable price less often than homogeneous markets.}
If agents cannot anticipate rivals' reaction functions, adjustment need not
settle. Under H3 the share of converged runs falls as the number of distinct
models in the market rises.

\hypothesis{H4 (Rent distribution)}{Inside a mixed market the collusive rent is shared unequally across models, and which model gains is not systematic across pairings.}
When two models settle at different prices, the lower-priced one captures
more demand. We have no prior on which model is the lower-priced one in a
given pairing; the null of interest is that the within-run profit gap between
the two models of a 2:2 cell is centered on zero, and the secondary question
is whether any ordering of models that emerges is transitive across the six
pairings.

H1 is a precondition for the rest: if models do not differ in homogeneous
markets, mixing them cannot matter. H2 and H3 are the natural predictions of
the focal-point account of tacit collusion. H4 asks who benefits when H2 or
H3 fails to hold, which turns out to be the informative case.

\section{Results}
\label{sec:results}

\begin{figure}[H]
\centering
\includegraphics[width=\textwidth]{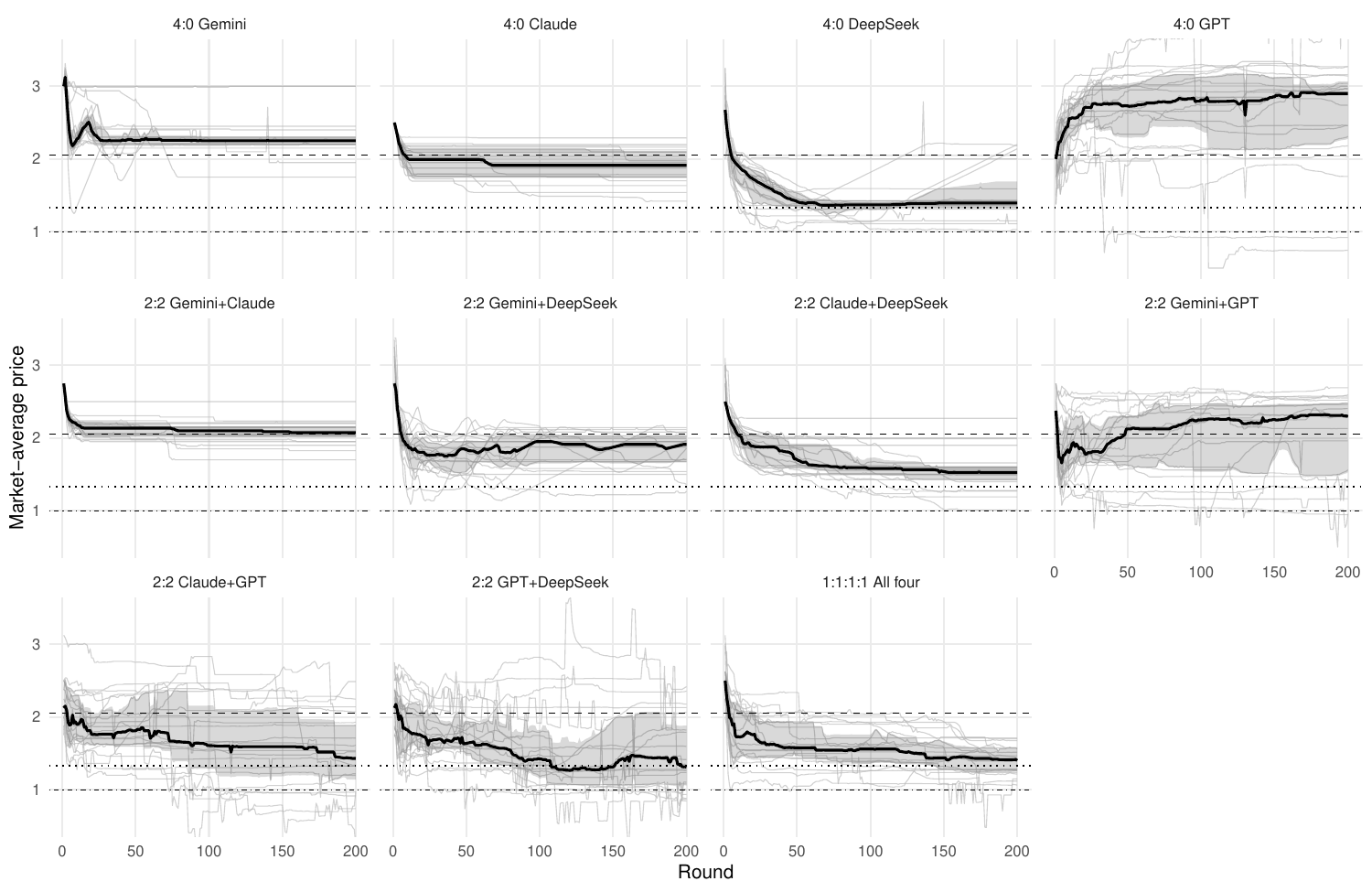}
\caption{Market-average price by period, all runs}
\label{fig:trajectories}
\begin{floatnote}[0.95\textwidth]
\emph{Note:} Gray lines are the 20 runs of a cell; the black line is the
period-wise median and the band the interquartile range. The dashed line is
the joint-monopoly price (2.05), the dotted line the Nash price (1.33), and
the dot-dashed line marginal cost (1.00).
\end{floatnote}
\end{figure}

Figure~\ref{fig:trajectories} shows the market-average price path in every
run of every cell. Two features of the trajectories organize everything that
follows. Runs that converge do so early, in most cells within the first 30
periods, and then hold a price that is typically identical across the four
firms to the cent. And the GPT panels look different in kind: prices drift
for the full 200 periods and never settle, whether GPT occupies all four
positions or only two.

\begin{figure}[H]
\centering
\includegraphics[width=\textwidth]{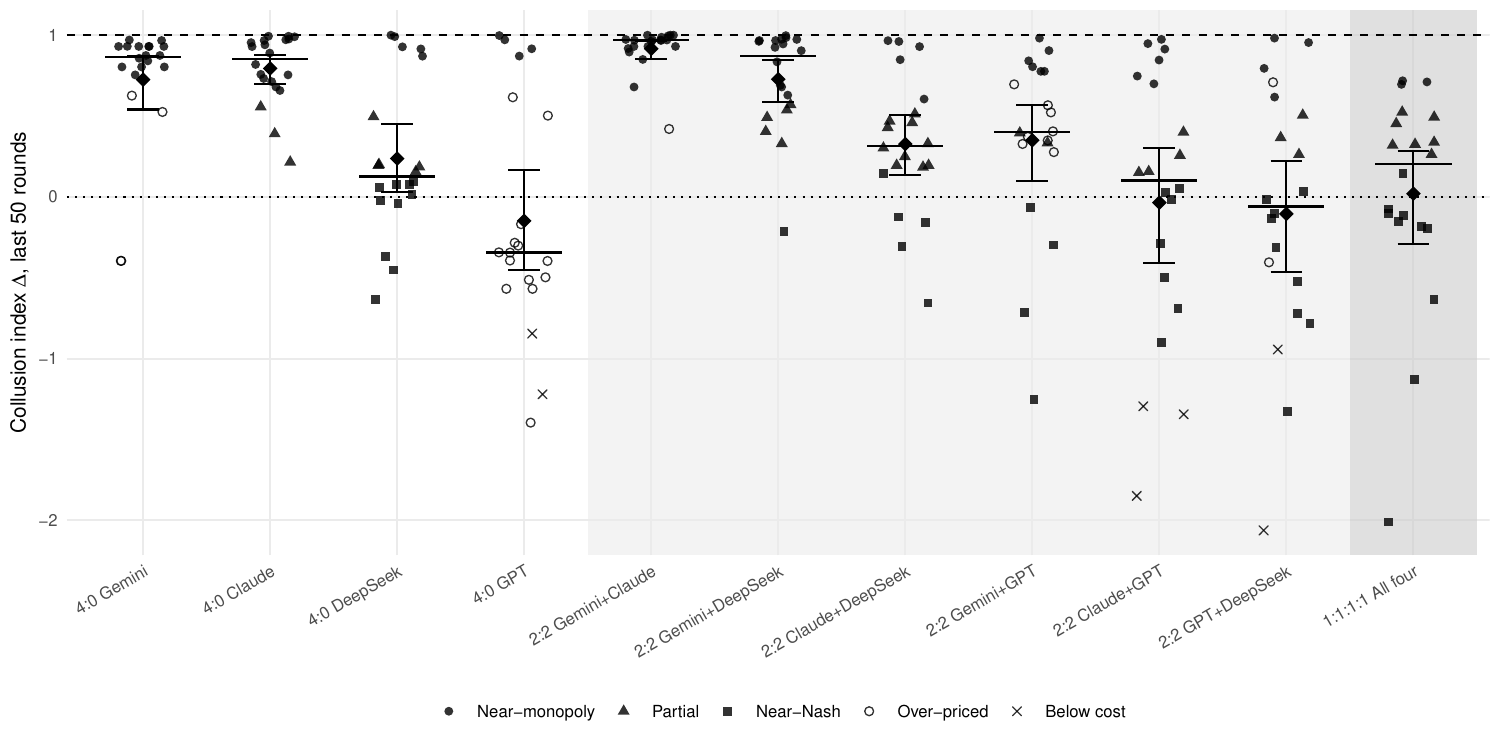}
\caption{Collusion index by cell, run level}
\label{fig:delta}
\begin{floatnote}[0.95\textwidth]
\emph{Note:} Each point is one run; the marker shape gives the outcome
category. The diamond
is the cell mean with a percentile-bootstrap 95\% interval; the horizontal
tick is the cell median. Shaded regions mark the 2:2:0:0 and 1:1:1:1
compositions.
\end{floatnote}
\end{figure}

Figure~\ref{fig:delta} collapses each run to its collusion index over the
last 50 periods and plots the 20 runs of every cell side by side. The
picture is not one of cells with a level and some noise around it. Outcomes
are multimodal within several cells, so that a cell mean can sit between
modes that no run occupies, and the modes are of different kinds: some runs
sit near the monopoly profit, some near Nash, and some have prices so far
above the monopoly level that profit is low for the opposite reason. This
is why the categorical outcome and the median are reported alongside the
mean throughout, and why all inference is non-parametric.

Table~\ref{tab:cells} gives the cell summaries that the rest of the section
draws on: the mean and median collusion index with a bootstrap interval, the
mean price and price index, the number of runs in each outcome category, and
the number of runs that converged. Four things are worth reading off the
table before the hypothesis tests. First, the eleven cells span almost the
full range of the index, from a mean $\Delta$ of 0.92 in the Gemini with
Claude cell to $-0.15$ in the homogeneous GPT cell, and the bootstrap
intervals of the highest and lowest cells do not come close to overlapping.
Whatever else is true, market composition matters. Second, the mean and the
median disagree in exactly the cells where the category counts show a
mixture. In the homogeneous Gemini cell the median is 0.87 and the mean 0.72
because four over-priced runs pull the mean down; in the GPT with DeepSeek
cell the median is $-0.06$ and the mean $-0.11$ because a few runs below
cost do the same. Where the category counts are concentrated, as in Gemini
with Claude (19 of 20 near-monopoly), mean and median coincide. Third, the
price index and the profit index tell different stories in the GPT cells.
Homogeneous GPT has the highest mean price in the table (2.62, a price index
of 1.78) and one of the lowest profit indices, and Gemini with GPT has a
price index above 1 with a profit index of 0.35. In these cells low $\Delta$
means prices above the monopoly level, not competition. Fourth, the
convergence column splits the table in two. Every cell without GPT converges
in at least 14 of 20 runs; no cell with GPT converges in more than 5. The
rest of this section takes these four observations in turn.

\begin{table}[H]
\centering
\caption{Cell outcomes, last 50 periods (20 runs per cell)}
\label{tab:cells}
\resizebox{\textwidth}{!}{
\begin{tabular}[t]{lrrcrrrrrrrr}
\toprule
\multicolumn{1}{c}{ } & \multicolumn{3}{c}{Collusion index $\Delta$} & \multicolumn{2}{c}{Price} & \multicolumn{5}{c}{Outcome category (runs)} & \multicolumn{1}{c}{ } \\
\cmidrule(l{3pt}r{3pt}){2-4} \cmidrule(l{3pt}r{3pt}){5-6} \cmidrule(l{3pt}r{3pt}){7-11}
Cell & Mean & Median & 95\% CI & $\bar p$ & $\Delta^{p}$ & Mono & Part & Nash & Over & $<$MC & Conv.\\
\midrule
4:0 Gemini & 0.72 & 0.87 & {}[0.54, 0.87] & 2.29 & 1.33 & 16 & 0 & 0 & 4 & 0 & 19\\
4:0 Claude & 0.79 & 0.85 & {}[0.70, 0.88] & 1.91 & 0.80 & 17 & 3 & 0 & 0 & 0 & 19\\
4:0 DeepSeek & 0.24 & 0.13 & {}[0.03, 0.45] & 1.51 & 0.24 & 5 & 5 & 10 & 0 & 0 & 14\\
4:0 GPT & -0.15 & -0.34 & {}[-0.45, 0.17] & 2.62 & 1.78 & 4 & 0 & 0 & 14 & 2 & 0\\
2:2 Gemini+Claude & 0.92 & 0.97 & {}[0.85, 0.96] & 2.08 & 1.03 & 19 & 0 & 0 & 1 & 0 & 20\\
2:2 Gemini+DeepSeek & 0.73 & 0.87 & {}[0.59, 0.85] & 1.82 & 0.68 & 14 & 5 & 1 & 0 & 0 & 14\\
2:2 Claude+DeepSeek & 0.33 & 0.32 & {}[0.14, 0.51] & 1.56 & 0.32 & 5 & 10 & 5 & 0 & 0 & 16\\
2:2 Gemini+GPT & 0.35 & 0.40 & {}[0.10, 0.57] & 2.06 & 1.01 & 6 & 2 & 4 & 8 & 0 & 1\\
2:2 Claude+GPT & -0.04 & 0.10 & {}[-0.41, 0.30] & 1.55 & 0.30 & 6 & 4 & 7 & 0 & 3 & 3\\
2:2 GPT+DeepSeek & -0.11 & -0.06 & {}[-0.46, 0.22] & 1.55 & 0.30 & 4 & 3 & 9 & 2 & 2 & 2\\
1:1:1:1 All four & 0.02 & 0.20 & {}[-0.29, 0.28] & 1.41 & 0.11 & 3 & 7 & 10 & 0 & 0 & 5\\
\bottomrule
\end{tabular}
}
\begin{floatnote}[0.95\textwidth]
\emph{Note:} $\Delta$ is the collusion index (\ref{eq:delta}); the interval
is a percentile bootstrap over runs. $\bar p$ is the market-average price and
$\Delta^{p}$ the price index. Categories are defined in
Section~\ref{sec:measures}. Conv.\ is the number of runs with at most five
price changes in the last 50 periods.
\end{floatnote}
\end{table}

\subsection{H1: Collusion is a property of the model}

The four homogeneous cells are not exchangeable (Kruskal-Wallis
$H = 23.99$, $p < 0.001$). They split into two groups
(Table~\ref{tab:pairwise}). Claude and Gemini markets reach mean $\Delta$ of
0.79 [0.70, 0.88] and 0.72 [0.54, 0.87]; the two are indistinguishable
($p = 0.62$, Cliff's $\delta = -0.10$). DeepSeek and GPT markets reach 0.24
[0.03, 0.45] and $-0.15$ [$-0.45$, 0.17]. Every comparison across the two
groups survives Holm correction with Cliff's $\delta$ between 0.50 and 0.69,
and the DeepSeek versus GPT comparison is marginal (Holm $p = 0.08$). In
run-level terms, 20 of 20 Claude runs and 18 of 20 Gemini runs earn
substantially more than Nash profit, against 10 of 20 for DeepSeek and 6 of
20 for GPT.

\begin{table}[H]
\centering
\caption{Pairwise comparisons of homogeneous cells}
\label{tab:pairwise}

\begin{tabular}[t]{lrrrrrr}
\toprule
Comparison & $\Delta$ diff.\ (mean) & (median) & $U$ & $p$ & Holm $p$ & Cliff's $\delta$\\
\midrule
Gemini vs.\ Claude & -0.07 & 0.01 & 181 & 0.616 & 0.616 & -0.10\\
Gemini vs.\ DeepSeek & 0.49 & 0.74 & 300 & 0.007 & 0.021 & 0.50\\
Gemini vs.\ GPT & 0.87 & 1.21 & 321 & 0.001 & 0.004 & 0.60\\
Claude vs.\ DeepSeek & 0.56 & 0.73 & 328 & $<$0.001 & 0.003 & 0.64\\
Claude vs.\ GPT & 0.94 & 1.20 & 339 & $<$0.001 & 0.001 & 0.69\\
DeepSeek vs.\ GPT & 0.39 & 0.47 & 277 & 0.039 & 0.077 & 0.39\\
\bottomrule
\end{tabular}

\begin{floatnote}[0.85\textwidth]
\emph{Note:} Mann-Whitney $U$ tests on run-level $\Delta$, 20 runs per cell,
Holm correction over the six comparisons. Cliff's $\delta$ is the probability
that a run from the first cell exceeds a run from the second minus the
reverse probability. Kruskal-Wallis across the four cells: $H = 23.99$,
$p < 0.001$.
\end{floatnote}
\end{table}

\begin{figure}[H]
\centering
\includegraphics[width=\textwidth]{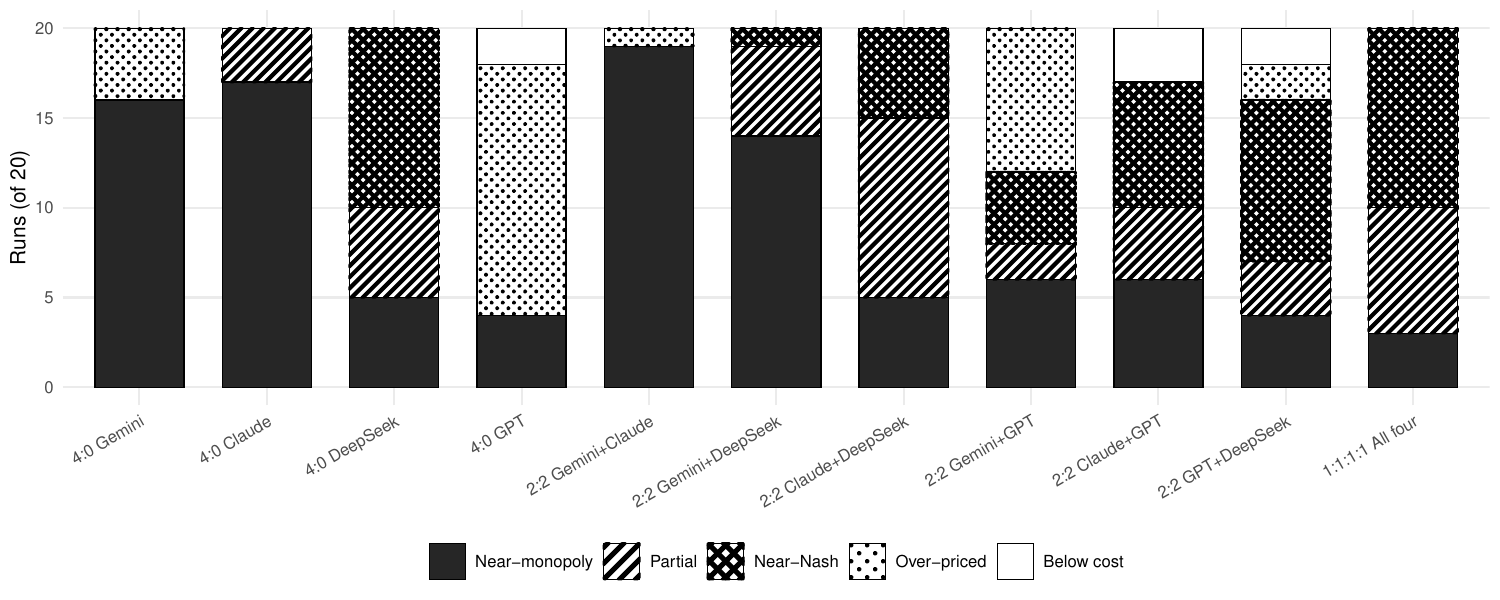}
\caption{Outcome categories by cell}
\label{fig:categories}
\end{figure}

The groups differ qualitatively as well as in level, and the categorical
breakdown in Table~\ref{tab:cells} and Figure~\ref{fig:categories} is more
informative than the means. Claude markets lock within 10 to 15 periods on a
single price between 1.75 and 2.25 and stay there; 17 of 20 runs are
near-monopoly and none is over-priced or below cost. Gemini markets also lock
early, but 4 of 20 runs freeze at 3.00, the price the agents happened to post
in the first period, which lies far above the monopoly price and leaves each
firm with less than half of Nash profit. The agents' own text describes this
state as a stable cooperative equilibrium from which no firm has an incentive
to deviate. It is a focal-point failure rather than competition, and it is
why the Gemini mean falls below its median of 0.87. DeepSeek markets are
bimodal: half of the runs settle near Nash (1.21 to 1.43) after a sequence of
one-cent undercuts, half settle between 1.6 and 2.2, and in both cases the
agents describe the resting point as tacit collusion. GPT markets do not
settle at all. Fourteen of 20 drift to between 2.8 and 3.2, above the
monopoly price, and two slide below marginal cost. The GPT agents' plans
contain no profit calculation, only a rule to stay near rivals and adjust if
the gap exceeds a threshold, so the market follows whichever firm last
moved.

Two consequences follow for how the rest of the results should be read. The
propensity to collude is a fixed characteristic of the model, and it varies
widely even within the cheapest tier of the market. And a negative $\Delta$
does not mean competition. The GPT cell's mean price of 2.62 is above the
monopoly price; consumers in that market pay more than they would under
collusion. We therefore report $\Delta$ together with $\Delta^{p}$ and the
category counts throughout.

\subsection{How each model prices}
\label{sec:mechanism}

Before turning to mixed markets we describe the pricing behavior behind
the differences across homogeneous cells, because the mixed-market results
depend on it. Table~\ref{tab:firstprice} reports opening prices
and Table~\ref{tab:mechanism} price adjustment by model and phase.

\begin{table}[H]
\centering
\caption{Opening price by model}
\label{tab:firstprice}

\begin{tabular}[t]{lrrr}
\toprule
Model & Mean & SD & Share above $p^M + 0.3$\\
\midrule
Gemini & 3.00 & 0.00 & 1.00\\
Claude & 2.50 & 0.00 & 1.00\\
GPT & 1.87 & 0.71 & 0.45\\
DeepSeek & 2.74 & 0.70 & 0.93\\
\bottomrule
\end{tabular}

\begin{floatnote}[0.7\textwidth]
\emph{Note:} Price posted in period 1, when the agent has no history. 220
firm-runs per model.
\end{floatnote}
\end{table}

Two of the four models open at a fixed price regardless of anything else.
Every Gemini firm in every run opens at 3.00 and every Claude firm at 2.50;
the standard deviation across 220 firm-runs is zero in both cases. DeepSeek
opens at 2.74 on average and GPT at 1.87, both with substantial dispersion.
Since 3.00 and 2.50 are both above the monopoly price, the first thing that
happens in a market containing Gemini or Claude is a descent.

\begin{table}[H]
\centering
\caption{Price adjustment by model and phase}
\label{tab:mechanism}
\resizebox{\textwidth}{!}{
\begin{tabular}[t]{llrrrrrc}
\toprule
\multicolumn{2}{c}{ } & \multicolumn{2}{c}{Price changes} & \multicolumn{3}{c}{Fresh undercut by a rival} & \multicolumn{1}{c}{Reaction to rival gap} \\
\cmidrule(l{3pt}r{3pt}){3-4} \cmidrule(l{3pt}r{3pt}){5-7} \cmidrule(l{3pt}r{3pt}){8-8}
Model & Phase & Share of periods with a change & Share of firm-runs never changing & Events & P(cut next period) & P(match) & Slope [95\% CI]\\
\midrule
Gemini & Periods 2--30 & 0.50 & 0.00 & 480 & 0.51 & 0.28 & 0.41 [0.34, 0.48]\\
 & Periods 31--200 & 0.08 & 0.33 & 477 & 0.22 & 0.48 & 0.05 [0.02, 0.08]\\
Claude & Periods 2--30 & 0.27 & 0.00 & 295 & 0.54 & 0.41 & 0.12 [0.08, 0.16]\\
 & Periods 31--200 & 0.03 & 0.41 & 228 & 0.18 & 0.59 & 0.02 [0.01, 0.03]\\
GPT & Periods 2--30 & 0.40 & 0.00 & 427 & 0.23 & 0.15 & 0.07 [0.06, 0.09]\\
 & Periods 31--200 & 0.23 & 0.00 & 1832 & 0.30 & 0.68 & 0.05 [0.04, 0.07]\\
DeepSeek & Periods 2--30 & 0.77 & 0.00 & 544 & 0.63 & 0.42 & 0.35 [0.29, 0.41]\\
 & Periods 31--200 & 0.27 & 0.08 & 1061 & 0.31 & 0.63 & 0.14 [0.09, 0.20]\\
\bottomrule
\end{tabular}
}
\begin{floatnote}[0.95\textwidth]
\emph{Note:} Firm-period observations pooled over all cells in which the
model appears. A change is a move of more than 0.005. A firm-run is
``never changing'' if the firm made no change in the phase. A fresh
undercut is the first period in which a rival is at least a cent below the
firm; P(cut) is the probability that the firm lowers its price in the next
period and P(match) that it moves to within two cents of that rival. The
slope is the coefficient from regressing the firm's price change on the
previous period's gap between its cheapest rival and itself; the interval is
a cluster bootstrap over runs.
\end{floatnote}
\end{table}

The descent is fast and then stops. In periods 2 to 30 Gemini changes its
price in half of all periods and DeepSeek in three quarters; both respond
strongly to the gap to their cheapest rival (slopes 0.41 and 0.35) and both
cut in the period after a fresh undercut more often than not (0.51 and
0.63). Claude moves less (27 percent of periods, slope 0.12) but also cuts
after roughly half of fresh undercuts. After period 30 Gemini and Claude
almost stop: they change price in 8 and 3 percent of periods, a third and
two fifths of their firm-runs never change again, their reaction slopes fall
to 0.05 and 0.02, and a fresh undercut now draws a cut only 22 and 18
percent of the time. DeepSeek keeps adjusting (27 percent of periods, cut
probability 0.31), and GPT never freezes at all: no GPT firm-run is without
a change after period 30, and GPT's dominant response to an undercut in the
late phase is to match it (0.68) rather than to cut below.

Two corrections to the intuitive reading of Figure~\ref{fig:trajectories}
follow. Gemini is not a model that posts a price and holds it; it adjusts as
actively as any model in the first 30 periods. What distinguishes it is
where it starts and where it stops: it opens highest and its descent ends
highest. Nor is Claude a model that follows rivals down indefinitely; it
follows for the first 30 periods and then, like Gemini, stops responding.
The behavior that matters for mixed markets is therefore a level effect
set in the opening phase, combined with a shared late-phase refusal to
retaliate.

\begin{table}[H]
\centering
\caption{Strategic vocabulary in plans and insights, by model}
\label{tab:keywords}

\begin{tabular}[t]{lrrrr}
\toprule
Term & Gemini & Claude & GPT & DeepSeek\\
\midrule
price war & 25.6 & 3.9 & 4.4 & 52.8\\
undercut & 13.5 & 13.4 & 34.0 & 67.9\\
match & 11.7 & 8.3 & 4.7 & 42.6\\
stable / stability & 77.2 & 83.5 & 29.0 & 62.0\\
collud / collusion & 23.2 & 20.8 & 0.1 & 33.9\\
cooperat / coordinat & 53.2 & 46.4 & 0.5 & 24.0\\
hold / maintain & 93.8 & 97.3 & 74.9 & 72.1\\
raise / increase & 16.1 & 3.8 & 10.5 & 13.3\\
lower / cut / reduce & 24.2 & 9.3 & 12.0 & 36.5\\
profit & 87.5 & 92.5 & 71.0 & 85.7\\
\bottomrule
\end{tabular}

\begin{floatnote}[0.7\textwidth]
\emph{Note:} Percentage of firm-periods (44{,}000 per model, pooled over
all cells) whose plan or insight text contains the term. Matching is
case-insensitive on the stem shown.
\end{floatnote}
\end{table}

The agents' own text is consistent with this reading, with the caveat that
stated reasons need not be actual ones \citep{turpin2023language}.
Table~\ref{tab:keywords} counts strategic terms in the plans and insights.
Gemini and Claude describe the market in the vocabulary of coordination:
``stable'' appears in 77 and 84 percent of their firm-periods,
``cooperate'' or ``coordinate'' in 53 and 46 percent, ``collude'' in 23 and
21 percent, and ``hold'' or ``maintain'' in over 90 percent. DeepSeek is the
only model that talks predominantly about conflict: ``undercut'' appears in
68 percent of its firm-periods, ``price war'' in 53 percent and ``match'' in
43 percent, against 4 to 34 percent for the others. GPT's text is
different in kind. ``Collude'' and ``cooperate'' appear in 0.1 and 0.5
percent of its firm-periods, ``stable'' in 29 percent, and ``profit'' in 71
percent against 86 to 93 percent for the other three. Whatever GPT is doing
when it drifts, it is not describing it as a strategic interaction.

\subsection{H2: Mixing raises collusion when the mix contains an anchor}

Table~\ref{tab:mixing} reports the mixing effect for each heterogeneous
cell. Pooling all mixed runs against all pure runs gives 0.31 against 0.40,
a difference that is not significant ($p = 0.38$, Cliff's $\delta = -0.07$).
H2 in its unconditional form is not supported. The cell-level estimates
show why: the sign of the effect depends on which model is in the mix.

\begin{table}[H]
\centering
\caption{Mixing effects}
\label{tab:mixing}
\resizebox{\textwidth}{!}{
\begin{tabular}[t]{lrrrcrrrr}
\toprule
\multicolumn{1}{c}{ } & \multicolumn{2}{c}{Mean $\Delta$} & \multicolumn{2}{c}{Mixing effect} & \multicolumn{2}{c}{Permutation} & \multicolumn{1}{c}{MW} & \multicolumn{1}{c}{Cliff's} \\
\cmidrule(l{3pt}r{3pt}){2-3} \cmidrule(l{3pt}r{3pt}){4-5} \cmidrule(l{3pt}r{3pt}){6-7} \cmidrule(l{3pt}r{3pt}){8-8} \cmidrule(l{3pt}r{3pt}){9-9}
Cell & Mixed & Pure avg. & Estimate & 95\% CI & $p$ & Holm $p$ & $p$ & $\delta$\\
\midrule
2:2 Gemini+Claude & 0.92 & 0.76 & +0.16 & {}[0.05, 0.27] & 0.035 & 0.211 & 0.003 & 0.47\\
2:2 Gemini+DeepSeek & 0.73 & 0.48 & +0.25 & {}[0.05, 0.43] & 0.049 & 0.245 & 0.053 & 0.31\\
2:2 Claude+DeepSeek & 0.33 & 0.52 & -0.19 & {}[-0.41, 0.03] & 0.131 & 0.393 & 0.122 & -0.25\\
2:2 Gemini+GPT & 0.35 & 0.29 & +0.06 & {}[-0.25, 0.35] & 0.740 & 0.873 & 0.535 & -0.10\\
2:2 Claude+GPT & -0.04 & 0.32 & -0.36 & {}[-0.76, 0.01] & 0.091 & 0.363 & 0.086 & -0.28\\
2:2 GPT+DeepSeek & -0.11 & 0.04 & -0.15 & {}[-0.55, 0.22] & 0.437 & 0.873 & 0.567 & -0.09\\
1:1:1:1 All four & 0.02 & 0.40 & -0.38 & {}[-0.71, -0.11] & 0.017 & 0.117 & 0.007 & -0.39\\
\midrule
All mixed vs.\ all pure & 0.31 & 0.40 & -0.09 & & & & 0.382 & -0.07 \\
\bottomrule
\end{tabular}
}
\begin{floatnote}[0.9\textwidth]
\emph{Note:} The mixing effect is the mixed cell's mean $\Delta$ minus the
average of its constituent homogeneous cells' means. The interval is a
percentile bootstrap over runs. The permutation $p$ reshuffles run labels
between the mixed cell and its constituent pure cells (20{,}000 draws). The
MW column tests the mixed cell against the pooled constituent pure runs.
\end{floatnote}
\end{table}

Markets that contain Gemini are more collusive than the homogeneous
markets they are built from. Gemini with Claude reaches 0.92 [0.85, 0.96],
the highest of the eleven cells and above either constituent (effect
$+0.16$ [0.05, 0.27], permutation $p = 0.04$). Gemini with DeepSeek lifts
the DeepSeek cell from 0.24 to 0.73 (effect $+0.25$ [0.05, 0.43],
$p = 0.05$), and inside that cell the DeepSeek firms themselves reach a
model-level $\Delta$ of 0.87 (Table~\ref{tab:within}), roughly what Claude
firms earn in a Claude market. Markets that contain Claude but not Gemini
are less collusive than their constituents: Claude with DeepSeek falls from
a constituent average of 0.52 to 0.33 (effect $-0.19$ [$-0.41$, 0.03],
$p = 0.13$), and Claude with GPT from 0.32 to $-0.04$ (effect $-0.36$
[$-0.76$, 0.01], $p = 0.09$). The fully heterogeneous 1:1:1:1 market is
significantly less collusive than the average homogeneous market (effect
$-0.38$ [$-0.71$, $-0.11$], $p = 0.02$; Mann-Whitney $p = 0.007$), with 10
of 20 runs ending near Nash.

Section~\ref{sec:mechanism} explains the pattern. Gemini opens at 3.00,
half a unit above Claude's 2.50 and above every other model's average, and
its descent stops after about 30 periods; because Gemini and Claude both
stop responding once the market has settled, whatever price the market has
reached by then is where it stays. A market that contains Gemini therefore
settles higher than one that does not, whoever the partner is, and the
partner, if it is a model that stops adjusting when its rivals do, settles
there with it. A market that contains Claude but not Gemini starts lower and
is pulled further down in the opening phase by DeepSeek, which cuts after
almost two thirds of fresh undercuts in that phase, before Claude's
late-phase inertia locks the result in. Gemini is an anchor in the sense
that it sets the level at which the market freezes, not in the sense that it
refuses to move. The fully mixed market contains one Gemini firm among
four, and its opening price is pulled down by three rivals who start lower
and one who never freezes; 10 of its 20 runs end near Nash. Heterogeneity
as such does not reduce collusion. What matters is whether the market
contains a model whose opening price and early descent end high, and whether
the remaining models stop adjusting when it does.

Three caveats apply to Table~\ref{tab:mixing}. None of the seven cell-level
effects survives Holm correction across the seven tests, so the evidence is
the pattern of signs and magnitudes rather than any single cell. The
batch-stratified permutation (Appendix Table~\ref{tab:mixbatch})
reproduces the same $p$-values, but the Gemini with GPT effect reverses
sign between batches ($-0.30$ and $+0.43$), and we do not interpret that
cell. The fully mixed cell is the exception on both counts: its effect is
significant on the Mann-Whitney test after Holm correction ($p = 0.049$)
and is nearly identical in the two batches ($-0.39$ and $-0.37$).

\begin{table}[H]
\centering
\caption{Model-level collusion index inside mixed cells}
\label{tab:within}

\begin{tabular}[t]{lcccc}
\toprule
Cell & Gemini & Claude & GPT & DeepSeek\\
\midrule
2:2 Gemini+Claude & 0.88 (2.09) & 0.95 (2.07) & -- & --\\
2:2 Gemini+DeepSeek & 0.59 (1.85) & -- & -- & 0.87 (1.79)\\
2:2 Claude+DeepSeek & -- & 0.27 (1.59) & -- & 0.39 (1.53)\\
2:2 Gemini+GPT & 0.52 (2.05) & -- & 0.17 (2.07) & --\\
2:2 Claude+GPT & -- & 0.25 (1.50) & -0.32 (1.60) & --\\
2:2 GPT+DeepSeek & -- & -- & -0.48 (1.63) & 0.26 (1.47)\\
1:1:1:1 All four & 0.09 (1.45) & 0.16 (1.41) & -0.29 (1.36) & 0.12 (1.42)\\
\bottomrule
\end{tabular}

\begin{floatnote}[0.75\textwidth]
\emph{Note:} Each entry is the mean $\Delta$ of the firms running that model
in that cell, with the model's mean price in parentheses.
\end{floatnote}
\end{table}

\subsection{H3: Instability is contagious, but it comes from one model}

\begin{figure}[H]
\centering
\includegraphics[width=\textwidth]{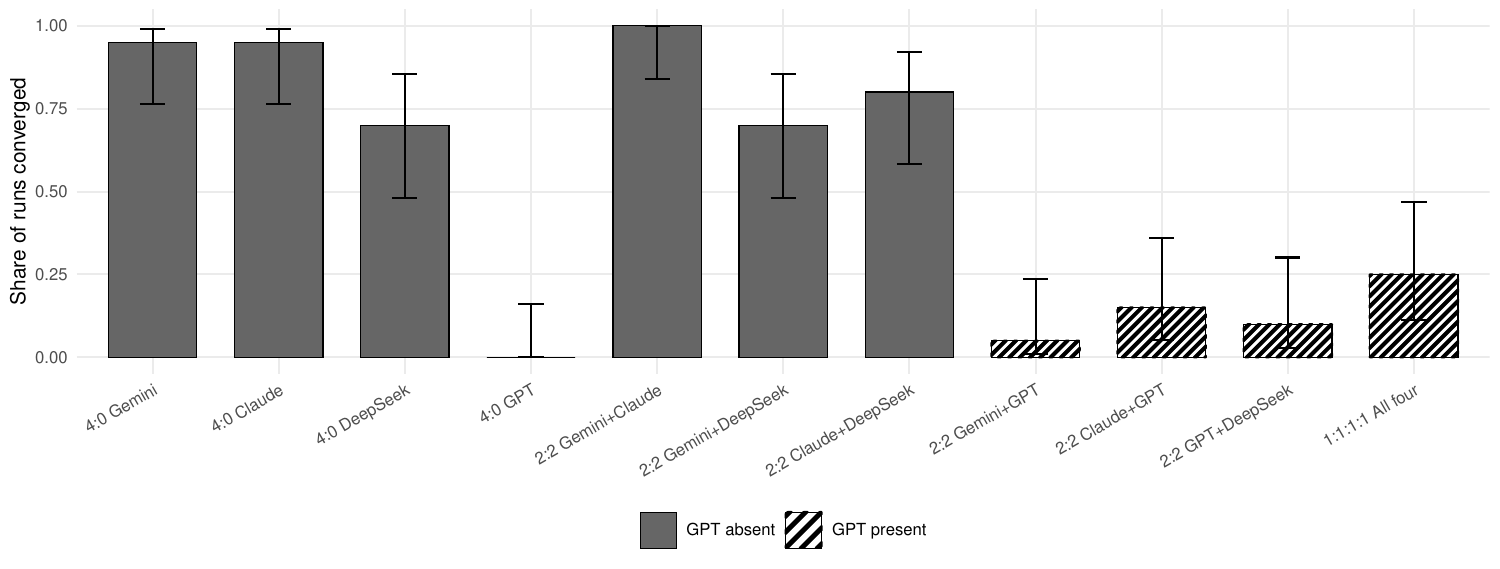}
\caption{Share of converged runs by cell}
\label{fig:convergence}
\end{figure}

H3 predicts that convergence falls with the number of distinct models. It
does not (Table~\ref{tab:convergence}, Figure~\ref{fig:convergence}). Among
the cells that do not contain GPT, 52 of 60 homogeneous runs and 50 of 60
two-by-two runs converge (Fisher $p = 0.80$). Gemini with Claude converges
in 20 of 20 runs, Claude with DeepSeek in 16, Gemini with DeepSeek in 14,
each within a few runs of its constituent pure cells. Heterogeneity by itself
does not destabilize the market.

\begin{table}[H]
\centering
\caption{Convergence by cell}
\label{tab:convergence}

\begin{tabular}[t]{lrrcrr}
\toprule
\multicolumn{1}{c}{ } & \multicolumn{3}{c}{Converged runs} & \multicolumn{2}{c}{Price changes, last 50 rounds} \\
\cmidrule(l{3pt}r{3pt}){2-4} \cmidrule(l{3pt}r{3pt}){5-6}
Cell & Count & Share & Wilson 95\% CI & Mean & Median\\
\midrule
4:0 Gemini & 19 & 0.95 & {}[0.76, 0.99] & 0.4 & 0\\
4:0 Claude & 19 & 0.95 & {}[0.76, 0.99] & 0.9 & 0\\
4:0 DeepSeek & 14 & 0.70 & {}[0.48, 0.85] & 8.2 & 1\\
4:0 GPT & 0 & 0.00 & {}[0.00, 0.16] & 22.8 & 22\\
2:2 Gemini+Claude & 20 & 1.00 & {}[0.84, 1.00] & 0.3 & 0\\
2:2 Gemini+DeepSeek & 14 & 0.70 & {}[0.48, 0.85] & 11.4 & 0\\
2:2 Claude+DeepSeek & 16 & 0.80 & {}[0.58, 0.92] & 3.0 & 0\\
2:2 Gemini+GPT & 1 & 0.05 & {}[0.01, 0.24] & 16.1 & 17\\
2:2 Claude+GPT & 3 & 0.15 & {}[0.05, 0.36] & 18.9 & 20\\
2:2 GPT+DeepSeek & 2 & 0.10 & {}[0.03, 0.30] & 27.4 & 32\\
1:1:1:1 All four & 5 & 0.25 & {}[0.11, 0.47] & 20.7 & 20\\
\bottomrule
\end{tabular}

\begin{floatnote}[0.8\textwidth]
\emph{Note:} A run has converged if at most five of its last 50 periods
contain a price change larger than 0.005 by any firm. Wilson intervals for
the share. Fisher's exact test, GPT present versus absent: 11 of 100 versus
102 of 120 converged, $p < 0.001$, odds ratio 0.02.
\end{floatnote}
\end{table}

What does destabilize it is the presence of GPT. Across the five cells that
contain GPT, 11 of 100 runs converge; across the six that do not, 102 of 120
(Fisher $p < 0.001$). GPT need not be in the majority. With two GPT firms
out of four, Gemini with GPT converges in 1 of 20 runs, Claude with GPT in
3, and DeepSeek with GPT in 2, so pairing GPT with the most stable model in
the study (Gemini, 19 of 20 in its own market) is enough to prevent the
market from settling. The mechanism is visible in the trajectories. The GPT
agents move a little toward their rivals' prices every period; the partner
model reads the movement as a market signal and responds; and the exchange
does not terminate. In the pilot, homogeneous GPT markets given 300 periods
still had not converged. Stability is thus not a function of how many
decision rules are present but of the least stable rule present, and one
such rule is enough to keep the whole market moving.

\subsection{H4: The anchor sets the price and the followers take the rent}

Table~\ref{tab:rent} examines who earns the collusive rent inside each 2:2
cell. For every run we compute the difference between the two models' mean
profit, expressed in units of $\pi^M - \pi^N$, and test it against zero with
the Wilcoxon signed-rank test. Ties are common because the two models often
lock on exactly the same price; the sign test therefore uses only the runs
in which one model is ahead.

\begin{table}[H]
\centering
\caption{Within-cell profit gap between the two models of a 2:2 market}
\label{tab:rent}
\resizebox{\textwidth}{!}{
\begin{tabular}[t]{lrrrcrcr}
\toprule
\multicolumn{1}{c}{ } & \multicolumn{2}{c}{Mean price} & \multicolumn{2}{c}{$\Delta$ gap (mean)} & \multicolumn{1}{c}{Wilcoxon} & \multicolumn{1}{c}{Ahead / tied /} & \multicolumn{1}{c}{Sign test} \\
\cmidrule(l{3pt}r{3pt}){2-3} \cmidrule(l{3pt}r{3pt}){4-5} \cmidrule(l{3pt}r{3pt}){6-6} \cmidrule(l{3pt}r{3pt}){7-7} \cmidrule(l{3pt}r{3pt}){8-8}
Pair (first $-$ second) & First & Second & Estimate & 95\% CI & $p$ & behind & $p$\\
\midrule
Gemini $-$ Claude & 2.09 & 2.07 & -0.07 & {}[-0.15, -0.02] & 0.021 & 0 / 14 / 6 & 0.031\\
Gemini $-$ DeepSeek & 1.85 & 1.79 & -0.28 & {}[-0.45, -0.13] & $<$0.001 & 1 / 5 / 14 & $<$0.001\\
Claude $-$ DeepSeek & 1.59 & 1.53 & -0.12 & {}[-0.21, -0.04] & 0.024 & 3 / 7 / 10 & 0.092\\
Gemini $-$ GPT & 2.05 & 2.07 & +0.35 & {}[0.13, 0.60] & 0.003 & 15 / 1 / 4 & 0.019\\
Claude $-$ GPT & 1.50 & 1.60 & +0.57 & {}[0.28, 0.91] & $<$0.001 & 18 / 1 / 1 & $<$0.001\\
GPT $-$ DeepSeek & 1.63 & 1.47 & -0.74 & {}[-1.15, -0.36] & 0.002 & 5 / 1 / 14 & 0.064\\
\bottomrule
\end{tabular}
}
\begin{floatnote}[0.95\textwidth]
\emph{Note:} The gap is the first model's mean profit minus the second's,
per run, in units of $\pi^M - \pi^N$; the interval is a percentile bootstrap
over the 20 runs. Wilcoxon signed-rank tests the gap against zero, dropping
exact ties. The next column counts runs in which the first model is ahead by
more than 0.005, tied, and behind; the sign test uses the non-tied runs.
Friedman test across the four model-level profits in the 1:1:1:1 cell:
$\chi^2 = 9.33$, $p = 0.03$.
\end{floatnote}
\end{table}

The gap is different from zero in all six pairings and its direction is
transitive. DeepSeek earns more than Claude (gap $-0.12$ for Claude,
$p = 0.02$), Claude earns more than Gemini ($-0.07$, $p = 0.02$; Gemini is
ahead in 0 of the 6 non-tied runs), DeepSeek earns more than Gemini
($-0.28$, $p < 0.001$; Gemini ahead in 1 of 15), and each of the three earns
more than GPT ($+0.35$, $+0.57$ and $+0.74$ in favor of Gemini, Claude and
DeepSeek respectively, all $p \le 0.003$). The ordering
DeepSeek $>$ Claude $>$ Gemini $\gg$ GPT contains no cycle, and the same
ordering appears in the 1:1:1:1 cell, where model-level $\Delta$ is 0.16
for Claude, 0.12 for DeepSeek, 0.09 for Gemini and $-0.29$ for GPT
(Friedman $p = 0.03$). H4's prediction of unequal sharing is supported; its
conjecture that the direction is unsystematic is rejected. The ordering is
also stable under resampling: in 99.95 percent of 10{,}000 bootstrap draws
that resample the 20 runs within every 2:2 cell, all six pairwise gaps keep
their sign.

\begin{figure}[H]
\centering
\includegraphics[width=\textwidth]{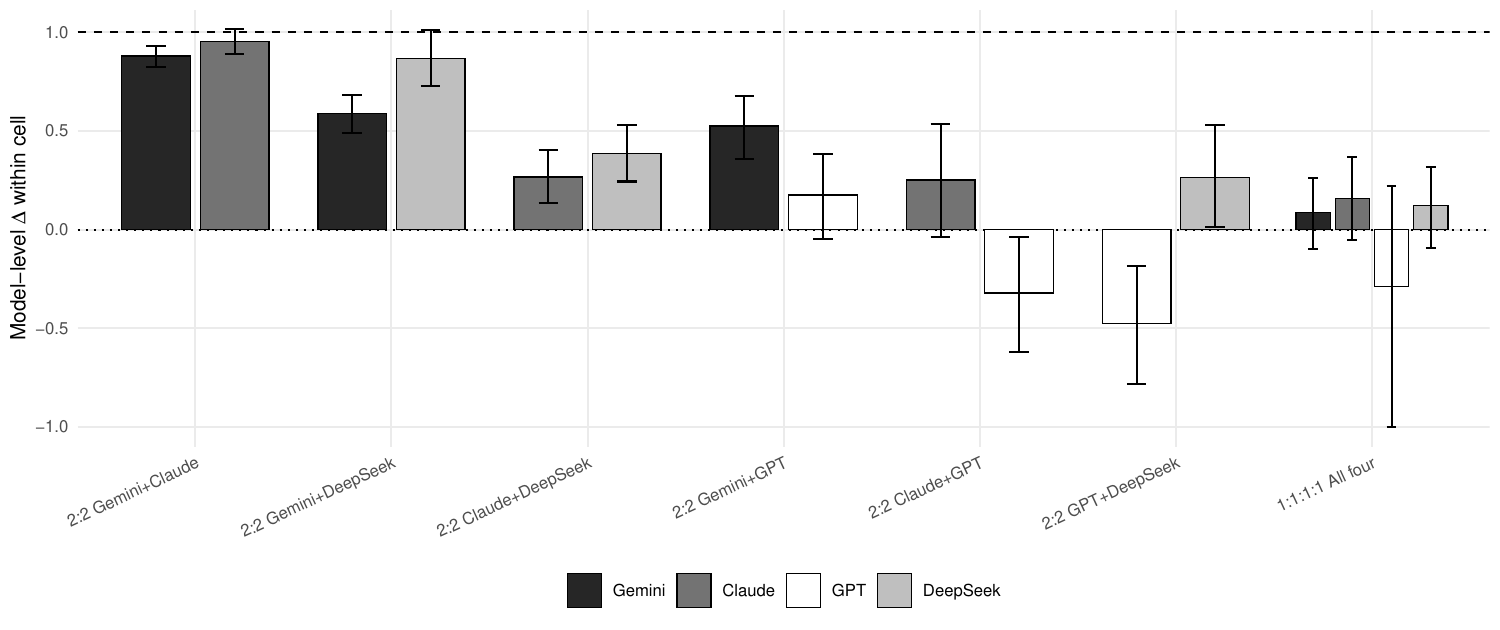}
\caption{Model-level collusion index inside mixed cells}
\label{fig:rent}
\begin{floatnote}[0.95\textwidth]
\emph{Note:} Bars are the mean $\Delta$ of the firms running each model in
each cell, with percentile-bootstrap 95\% intervals over runs.
\end{floatnote}
\end{figure}

The ordering inverts the anchor ranking of H2. Gemini is the model that
raises the market price, and it is the model that earns least among the
three collusive ones. In the Gemini with DeepSeek cell the Gemini firms hold
1.85 while the DeepSeek firms sit at 1.79, six cents below, and the six cents
are enough to transfer a quarter of the monopoly rent from Gemini to
DeepSeek in every period (Table~\ref{tab:within}). That a firm priced a few
cents below its rival earns more is, near the collusive price, a property
of logit demand rather than a finding. The finding is that the gap
persists. After period 30 a Gemini firm that has just been undercut cuts its
own price in the next period with probability 0.22, and a Claude firm with
probability 0.18 (Table~\ref{tab:mechanism}); DeepSeek, at 0.31, is the
model most willing to answer a cut with a cut and the model that ends up
below its partners. The leader does not retaliate, so the transfer persists
for the remaining 150 periods. GPT loses to every partner even though its
mean price is close to theirs, because its drift places it above the
partner when the partner has just cut and below when the partner has just
raised, so its timing is systematically wrong. This is the textbook
prediction of the price-leadership model: the leader supplies the high price
and absorbs the residual demand, while followers price just under it and
take the volume. Language-model markets reproduce it without any of the
agents being told who leads.

\subsection{Consumer surplus}

The collusion index measures the firms' side of the market. Because several
cells reach low profit through prices above the monopoly level rather than
below the Nash level, the consumer side does not follow from it.
Table~\ref{tab:welfare} reports consumer surplus over the outcome window,
normalized so that 0 is the joint-monopoly outcome and 1 is Nash.

\begin{table}[H]
\centering
\caption{Consumer surplus and total surplus by cell}
\label{tab:welfare}
\resizebox{\textwidth}{!}{
\begin{tabular}[t]{lrrcrr}
\toprule
\multicolumn{1}{c}{ } & \multicolumn{3}{c}{Consumer surplus} & \multicolumn{2}{c}{ } \\
\cmidrule(l{3pt}r{3pt}){2-4}
Cell & Level & Index & 95\% CI & Industry profit & Total-surplus index\\
\midrule
4:0 Gemini & 0.235 & -0.19 & {}[-0.27, -0.10] & 0.673 & -1.42\\
4:0 Claude & 0.499 & 0.21 & {}[0.08, 0.35] & 0.706 & 0.22\\
4:0 DeepSeek & 0.861 & 0.76 & {}[0.56, 0.95] & 0.439 & 0.75\\
4:0 GPT & 0.290 & -0.11 & {}[-0.38, 0.23] & 0.255 & -3.42\\
2:2 Gemini+Claude & 0.353 & -0.01 & {}[-0.09, 0.07] & 0.764 & -0.26\\
2:2 Gemini+DeepSeek & 0.573 & 0.32 & {}[0.19, 0.47] & 0.674 & 0.46\\
2:2 Claude+DeepSeek & 0.813 & 0.69 & {}[0.51, 0.86] & 0.482 & 0.73\\
2:2 Gemini+GPT & 0.458 & 0.15 & {}[-0.13, 0.47] & 0.493 & -1.18\\
2:2 Claude+GPT & 0.918 & 0.85 & {}[0.55, 1.15] & 0.309 & 0.35\\
2:2 GPT+DeepSeek & 0.963 & 0.91 & {}[0.63, 1.20] & 0.275 & 0.41\\
1:1:1:1 All four & 0.978 & 0.94 & {}[0.77, 1.13] & 0.335 & 0.82\\
\bottomrule
\end{tabular}
}
\begin{floatnote}[0.85\textwidth]
\emph{Note:} Consumer surplus is the logit inclusive value
$\mu \log(\sum_j e^{(a - p_j)/\mu} + e^{a_0/\mu})$, averaged over the last 50
periods and over runs; its level is 0.36 at the joint-monopoly price and
1.02 at the Nash price. The consumer-surplus index is $(CS - CS^M)/(CS^N -
CS^M)$ with a percentile-bootstrap interval over runs; the total-surplus
index normalizes consumer surplus plus industry profit the same way.
Negative values are outcomes worse than joint monopoly.
\end{floatnote}
\end{table}

Three cells leave consumers worse off than a cartel would. The homogeneous
Gemini market has a consumer-surplus index of $-0.19$ [$-0.27$, $-0.10$]:
the four runs frozen at 3.00 cost consumers more than the sixteen collusive
runs gain them relative to monopoly. The homogeneous GPT market is at
$-0.11$, and Gemini with Claude, the most collusive cell by the profit
index, is at $-0.01$, indistinguishable from monopoly. At the other end the
fully mixed market (0.94), GPT with DeepSeek (0.91) and Claude with GPT
(0.85) deliver nearly the Nash level of consumer surplus, but not because
firms compete: in the GPT cells they do so through drift that also destroys
producer surplus, so that total surplus in the homogeneous GPT market is
$-3.42$ on the normalized scale, more than three monopoly-to-Nash gaps
below the cartel outcome. The two indices together separate three kinds of
market. Gemini and Claude markets transfer surplus from consumers to firms
in the way collusion does. DeepSeek markets and the fully mixed market
leave most of it with consumers. GPT markets destroy it.

\subsection{Summary}

Homogeneous markets tell us which models are collusive (H1). Mixed markets
tell us that the effect of heterogeneity runs through the presence of an
anchor rather than through heterogeneity itself (H2), that stability is
determined by the least stable model rather than by the number of models
(H3), and that the rent created by the anchor accrues to the models that
follow it (H4). The last two findings have a common structure: a single
model determines the market-level property, whether it is the anchor
setting the price or the drifter preventing convergence, and the other
models' identities determine how the consequences are shared.

\section{Conclusion}
\label{sec:conclusion}

We have asked what happens to algorithmic collusion when the algorithms are
not the same. Four language models from four providers, each at its
cheapest tier, priced in a four-firm logit Bertrand market in every
homogeneous, two-by-two and fully mixed composition. The answer is not that
heterogeneity reduces collusion, nor that it leaves collusion unchanged. It
is that the market-level properties studied in the algorithmic-collusion
literature, the level of prices and their stability, are each determined by
a single participant, and that the distribution of the resulting rent is
determined by the others.

Collusion is a property of the model: two of the four models reach
three-quarters of the monopoly rent in homogeneous markets, one reaches a
quarter, and one does not settle at all. Mixing them does not average these
dispositions. A market that contains Gemini, which opens at the highest
price and settles highest, is more collusive than the homogeneous markets
it is built from; a market that contains Claude, which opens lower and
follows its rivals down, is less so;
and only the fully mixed market, in which no model can anchor, is
significantly less collusive than the homogeneous average. Stability is set
by the least stable model: two GPT firms out of four keep any market from
converging. And inside mixed markets the rent flows away from the anchor,
in a transitive order that inverts the anchor ranking, exactly as the
theory of price leadership predicts.

Three implications follow. For the design of experiments on algorithmic
collusion, results from homogeneous markets do not transfer to
heterogeneous ones by averaging, and the composition of a market must be a
treatment in its own right. For the measurement of collusion, a profit index
alone misclassifies markets in which prices have drifted above the monopoly
level as competitive, and a price index and a categorical outcome are
needed alongside it. For policy, the firm whose algorithm raises the price
is not the firm that profits from it, so liability regimes that look for
the firm that led the price will find the one that gained least, and
regimes that look for the firm that gained most will find a follower that
never raised its price at all.

The design has limits that suggest next steps. We use one prompt, one
market and the cheapest tier of each model; \citet{fish2024algorithmic}
show that prompt wording matters, and larger models may behave differently.
The agents' plans and insights are descriptive text, not evidence of
mechanism. And the four models are a snapshot of a market for foundation
models that changes every few months, so the specific ranking we report
will not last, even if the structure of the result, one model anchoring,
one model destabilizing, and the followers collecting, does.

\clearpage
\appendix
\renewcommand{\thetable}{A\arabic{table}}
\renewcommand{\thefigure}{A\arabic{figure}}
\setcounter{table}{0}
\setcounter{figure}{0}
\section{Robustness}

The 20 runs of each cell were collected in two batches of 10 under identical
settings, the first between 13:25 and 16:17 KST on 16 September 2026 and the
second between 08:54 and 11:43 on 17 September 2026. Because the models are
served through provider APIs, we cannot rule out that a provider changed a
model between the batches. Table~\ref{tab:batches} compares the batches cell
by cell and Table~\ref{tab:mixbatch} recomputes the mixing effects within
each batch.

\begin{table}[H]
\centering
\caption{Consistency across the two batches of 10 runs}
\label{tab:batches}

\begin{tabular}[t]{lrrr}
\toprule
Cell & Batch 1 mean $\Delta$ & Batch 2 mean $\Delta$ & MW $p$\\
\midrule
4:0 Gemini & 0.84 & 0.61 & 0.760\\
4:0 Claude & 0.75 & 0.84 & 0.678\\
4:0 DeepSeek & 0.04 & 0.43 & 0.096\\
4:0 GPT & -0.14 & -0.15 & 0.850\\
2:2 Gemini+Claude & 0.89 & 0.94 & 0.677\\
2:2 Gemini+DeepSeek & 0.78 & 0.68 & 0.910\\
2:2 Claude+DeepSeek & 0.25 & 0.40 & 0.450\\
2:2 Gemini+GPT & 0.05 & 0.65 & 0.011\\
2:2 Claude+GPT & 0.04 & -0.11 & 0.678\\
2:2 GPT+DeepSeek & -0.14 & -0.07 & 0.970\\
1:1:1:1 All four & -0.02 & 0.06 & 0.623\\
\bottomrule
\end{tabular}

\begin{floatnote}[0.75\textwidth]
\emph{Note:} Mann-Whitney $U$ test of run-level $\Delta$ between the two
batches, 10 runs each. One of eleven comparisons (Gemini with GPT) is
significant at 5\%, and a second (homogeneous DeepSeek) is at 10\%.
\end{floatnote}
\end{table}

\begin{table}[H]
\centering
\caption{Mixing effects within each batch}
\label{tab:mixbatch}

\begin{tabular}[t]{lrrrr}
\toprule
\multicolumn{1}{c}{ } & \multicolumn{3}{c}{Mixing effect (mixed $-$ pooled pure)} & \multicolumn{1}{c}{ } \\
\cmidrule(l{3pt}r{3pt}){2-4}
Cell & Batch 1 & Batch 2 & Pooled & Stratified permutation $p$\\
\midrule
2:2 Gemini+Claude & +0.09 & +0.22 & +0.16 & 0.032\\
2:2 Gemini+DeepSeek & +0.34 & +0.16 & +0.25 & 0.051\\
2:2 Claude+DeepSeek & -0.15 & -0.23 & -0.19 & 0.129\\
2:2 Gemini+GPT & -0.30 & +0.43 & +0.06 & 0.743\\
2:2 Claude+GPT & -0.27 & -0.45 & -0.36 & 0.091\\
2:2 GPT+DeepSeek & -0.09 & -0.21 & -0.15 & 0.438\\
1:1:1:1 All four & -0.39 & -0.37 & -0.38 & 0.016\\
\bottomrule
\end{tabular}

\begin{floatnote}[0.8\textwidth]
\emph{Note:} The mixing effect within a batch is the mixed cell's mean
$\Delta$ minus the mean of the pooled constituent pure runs from the same
batch. The stratified permutation reshuffles run labels within batch only
(20{,}000 draws).
\end{floatnote}
\end{table}

Table~\ref{tab:robust} varies the outcome window and the convergence
threshold. Using the last 100 periods instead of the last 50 changes no cell
mean by more than 0.13, and the ranking of cells is unchanged except that
Claude with DeepSeek and Gemini with GPT, which differ by 0.02, swap
places. Counting a run as converged with at most 3, 5 or 10 price changes
in the last 50 periods, or at most 10 in the last 100, moves individual
cells by at most four runs and leaves the split between cells with and
without GPT intact.

\begin{table}[H]
\centering
\caption{Robustness to the outcome window and the convergence threshold}
\label{tab:robust}
\resizebox{\textwidth}{!}{
\begin{tabular}[t]{lrrcrrrr}
\toprule
\multicolumn{1}{c}{ } & \multicolumn{3}{c}{Mean $\Delta$} & \multicolumn{4}{c}{Converged runs, price changes} \\
\cmidrule(l{3pt}r{3pt}){2-4} \cmidrule(l{3pt}r{3pt}){5-8}
Cell & Last 50 & Last 100 & 95\% CI (last 100) & $\le 3$ & $\le 5$ & $\le 10$ & $\le 10$ of last 100\\
\midrule
4:0 Gemini & 0.72 & 0.72 & {}[0.53, 0.87] & 19 & 19 & 20 & 20\\
4:0 Claude & 0.79 & 0.80 & {}[0.70, 0.88] & 18 & 19 & 19 & 19\\
4:0 DeepSeek & 0.24 & 0.20 & {}[0.01, 0.39] & 14 & 14 & 15 & 13\\
4:0 GPT & -0.15 & -0.13 & {}[-0.42, 0.17] & 0 & 0 & 1 & 0\\
2:2 Gemini+Claude & 0.92 & 0.92 & {}[0.85, 0.96] & 19 & 20 & 20 & 20\\
2:2 Gemini+DeepSeek & 0.73 & 0.73 & {}[0.59, 0.85] & 14 & 14 & 15 & 12\\
2:2 Claude+DeepSeek & 0.33 & 0.36 & {}[0.19, 0.53] & 16 & 16 & 17 & 13\\
2:2 Gemini+GPT & 0.35 & 0.35 & {}[0.12, 0.56] & 0 & 1 & 4 & 0\\
2:2 Claude+GPT & -0.04 & -0.00 & {}[-0.37, 0.33] & 2 & 3 & 4 & 2\\
2:2 GPT+DeepSeek & -0.11 & -0.15 & {}[-0.48, 0.15] & 2 & 2 & 4 & 1\\
1:1:1:1 All four & 0.02 & 0.15 & {}[-0.08, 0.37] & 3 & 5 & 7 & 1\\
\bottomrule
\end{tabular}
}
\end{table}

Table~\ref{tab:parse} reports the frequency with which an agent's reply
could not be parsed. When a reply is unparseable the script keeps the firm's
previous price. This happened in 62 of 44{,}000 GPT firm-periods (0.14
percent), in 2 DeepSeek firm-periods and never for Gemini or Claude; it
cannot account for the GPT cells' behavior.

\begin{table}[H]
\centering
\caption{Unparsed replies by model}
\label{tab:parse}

\begin{tabular}[t]{lrrrr}
\toprule
Model & Firm-periods & Unparsed replies & Share (\%) & Runs affected\\
\midrule
Gemini & 44,000 & 0 & 0.00 & 0\\
Claude & 44,000 & 0 & 0.00 & 0\\
GPT & 44,000 & 62 & 0.14 & 45\\
DeepSeek & 44,000 & 2 & 0.00 & 2\\
\bottomrule
\end{tabular}

\end{table}

Table~\ref{tab:icc} justifies treating the run as the unit of observation.
For every run the collusion index is computed on three non-overlapping
50-period segments; the intraclass correlation measures how much of the
variation across segments is between runs rather than within them, and the
Spearman correlation shows how well the outcome at periods 51 to 100 already
predicts the outcome at periods 151 to 200.

\begin{table}[H]
\centering
\caption{Within-run dependence of the collusion index}
\label{tab:icc}
\resizebox{\textwidth}{!}{
\begin{tabular}[t]{lrrrr}
\toprule
Cell & Between-run var. & Within-run var. & ICC & Spearman $\rho$ (51--100 vs.\ 151--200)\\
\midrule
4:0 Gemini & 0.149 & 0.001 & 0.99 & 0.81\\
4:0 Claude & 0.036 & 0.001 & 0.97 & 0.98\\
4:0 DeepSeek & 0.161 & 0.021 & 0.89 & 0.79\\
4:0 GPT & 0.416 & 0.017 & 0.96 & 0.96\\
2:2 Gemini+Claude & 0.017 & 0.000 & 0.99 & 0.98\\
2:2 Gemini+DeepSeek & 0.078 & 0.018 & 0.81 & 0.59\\
2:2 Claude+DeepSeek & 0.118 & 0.024 & 0.83 & 0.70\\
2:2 Gemini+GPT & 0.244 & 0.035 & 0.88 & 0.54\\
2:2 Claude+GPT & 0.491 & 0.082 & 0.86 & 0.71\\
2:2 GPT+DeepSeek & 0.348 & 0.112 & 0.76 & 0.63\\
1:1:1:1 All four & 0.223 & 0.075 & 0.75 & 0.77\\
\bottomrule
\end{tabular}
}
\begin{floatnote}[0.95\textwidth]
\emph{Note:} $\Delta$ is computed on three non-overlapping 50-period segments
of each run (periods 51--100, 101--150, 151--200). The ICC is the share of
total variance across segments that is between runs. The last column is the
Spearman correlation across runs between $\Delta$ on periods 51--100 and on
periods 151--200.
\end{floatnote}
\end{table}

\clearpage
\bibliographystyle{apacite7}
\bibliography{references}

\end{document}